\documentclass[openacc]{rstransa}

\newcommand{\be}{\begin{equation}}
\newcommand{\ee}{\end{equation}}
\newcommand{\ba}{\begin{align}}
\newcommand{\ea}{\end{align}}
\newcommand{\nn}{\nonumber}
\newcommand{\K}{{\mathcal K}}
\newcommand{\tr}{\mathrm{tr}}
\newcommand{\diag}{\text{ diag }}

\usepackage{todonotes}
\usepackage{braket}
\usepackage[inline, shortlabels]{enumitem}
\newlist{enumi}{enumerate*}{1} 
\setlist[enumi]{itemsep = 1.125in, label=(\roman*)}

\begin{document}

\title{Quantum fermions from classical bits}

\author{
Christof Wetterich}

\address{Institut  f\"ur Theoretische Physik\\
Universit\"at Heidelberg\\
Philosophenweg 16, D-69120 Heidelberg}

\subject{Quantum field theory, fermions, cellular automata}

\keywords{Probabilistic cellular automata, quantum mechanics from classical statistics, generalized Thirring model for fermions}

\corres{Christof Wetterich\\
\email{C.Wetterich@ThPhys.Uni-Heidelberg.DE}}

\begin{abstract}
A simple probabilistic cellular automaton is shown to be equivalent to a relativistic fermionic quantum field theory with interactions. Occupation numbers for fermions are classical bits or Ising spins. The automaton acts deterministically on bit configurations. The genuinely probabilistic character of quantum physics is realized by probabilistic initial conditions. In turn, the probabilistic automaton is equivalent to the classical statistical system of a generalized Ising model. For a description of the probabilistic information at any given time quantum concepts as wave functions and non-commuting operators for observables emerge naturally. Quantum mechanics can be understood as a particular case of classical statistics. This offers prospects  to realize aspects of quantum computing in the form of probabilistic classical computing.
\end{abstract}


\begin{fmtext}
\section{Introduction\label{sec: introduction}}

The central aim of this article is a demonstration that quantum mechanics emerges as a particular case of classical statistical systems. No fundamental concepts beyond the classical statistical probabilities are needed. We describe a probabilistic cellular automaton which is completely equivalent to a fermionic quantum field theory with interactions. There is a one to one mapping of all expectation values of observables, such that the cellular automaton and the fermionic quantum field theory cannot be distinguished by any observation. Both describe the same physical reality. The cellular automaton acts at each discrete time step on configurations of Ising spins in a deterministic way, changing every given configuration into precisely one new configuration. The probabilistic aspects arise from probabilistic initial conditions, as characterized by a probability distribution over initial spin configurations. Such a probabilistic cellular automaton can be viewed as a generalized (classical) Ising model, and therefore as a 
\end{fmtext}
\maketitle
\noindent particular case of a classical statistical system. A quantum field theory for fermions is a quantum system. Our example therefore demonstrates that this particular quantum system is realized by a particular classical statistical system. Quantum mechanics emerges from classical statistics~\cite{CWEM, CWQMCS}. Incidentally, our example shows that there cannot be a no-go theorem preventing the embedding of quantum mechanics in classical statistics. A central assumption of existing no go theorems based on Bell's inequalities~\cite{BELL}, namely that classical correlation functions describe all correlations of ideal measurements, is not valid~\cite{Wetterich:2020kqi}. 

The equivalence of a probabilistic cellular automaton and a fermionic quantum field theory is based on four keys ingredients:
\begin{enumerate}[(1) , itemindent=26pt, leftmargin=0pt, itemsep=0pt, topsep=0pt, listparindent=11pt]
\item Fermions are Ising spins. The states of a system of fermions can be characterized by occupation numbers for every possible position or momentum, plus possible internal properties as spin or other quantities. These occupation numbers can take values one or zero, just as classical bits. Occupation numbers $n=(1,0)$ are directly associated to Ising spins $s=(1, -1)$ by $s=2n-1$. The direct equivalence of configurations of Ising spins with the occupation number basis for fermionic quantum systems has led to several proposals of a fermionic description for two-dimensional Ising-type models~\cite{PLECH, BER1, BER2, SAM, ITS, PLE1} within classical statistics. Within quantum field theories, it is the basis for several concepts of bosonization or fermionization~\cite{FUR, NAO, COL, DNS}. 

A general bit-fermion map between fermions and Ising spins for arbitrary systems and arbitrary dimension is based on identical evolution operators for both systems~\cite{CWFCS, CWFGI}. We employ here this map for a demonstration of equivalence~\cite{Wetterich:2020kqi, CWPCA} of a probabilistic cellular automaton with a fermionic quantum field theory that is a type of Thirring model~\cite{THI,KLA, AAR,FAIV}. In our case this is a map between a classical statistical and a quantum system, in contrast to other maps that remain either within the setting of classical statistics, or the setting of quantum field theories. Both the generalized Ising models and the fermionic quantum field theory are described by a classical statistical "overall probability distribution" for the possible configurations of Ising spins or occupation numbers at all times~\cite{Wetterich:2020kqi}. The quantum concepts as wave function or density matrix describe the probabilistic information for a time-local subsystem of configurations at a given time~\cite{CWPT, Wetterich:2020kqi}. They apply both to general classical statistical systems and to particular quantum systems. 
"Classical wave functions"~\cite{CWQPCS} are useful concepts in classical statistics, as in our case for the understanding of evolution for probabilistic cellular automata.
\label{Intro_1}
\item A continuous unitary evolution can lead at discrete time steps $t=t_{\textup{in}}+m\varepsilon$, $m\in \mathbb{N}$ to evolution operators that map each configuration of Ising spins to precisely one other configuration. This is the deterministic evolution of a (classical) cellular automaton~\cite{WOLF}. (We do not describe here the concept of quantum cellular automata~\cite{Arrighi2019}). This feature of discrete time steps is at the basis of t'Hooft's attempts to understand quantum mechanics from cellular automata in a deterministic way~\cite{GTH, ELZE}.
\label{Intro_2}
\item Quantum mechanics is in our view genuinely probabilistic. We therefore investigate probabilistic cellular automata for which the initial (and possibly also final) spin configurations occur with certain probabilities. The initial probability distribution at $t_{\textup{in}}$ corresponds to a boundary term for the quantum field theory. One typically wants to understand how the probabilistic information at $t_{\textup{in}}$ p  of observables at $t$ can be computed from the time-local probabilistic information at $t$. This probabilistic information is encoded in the density matrix (or wave function for pure states)~\cite{Wetterich:2020kqi}. The density matrix contains the time-local probabilities as diagonal elements. It also contains additional time-local probabilistic information, which can be used for the computation of time derivatives of observables or observables at times $t'$ in the neighborhood of $t$. At this point the particle-wave duality characteristic for quantum mechanics emerges. The particle aspect concerns the discreteness of the possible values for occupation numbers. The wave aspect arises from the continuous probability distribution, or more generally density matrix and wave function~\cite{CWIT, CWQF}.
\label{Intro_3}
\item The continuum limit leads to important simplifications and new quantum features. For a sufficiently smooth wave function or  density matrix for a probabilistic cellular automaton the discreteness of the time steps plays no longer a role if time differences are much larger than the step size $\varepsilon$. In this case the probabilistic cellular automaton can be described by a continuous time evolution. Imagining a step size of the order of the Planck time and a discretization of space with lattice distance of the Planck length, the continuum limit is realized for all practical purposes. The probabilistic character of the automaton is crucial since smoothness concerns the time-local probabilistic information. (The wave function for a single deterministic state is a $\delta $-function and therefore never smooth.) The time evolution of a cellular automaton is guaranteed to be unitary. This extends to the continuum limit. The unitary evolution becomes particularly apparent if one introduces a complex structure associated to particle-hole transformations or time reversal. In the continuum limit the time evolution becomes genuinely probabilistic. Since differences of the order~$\varepsilon$ in time or space are no longer resolved, any given configurations evolves in a probabilistic way to a whole set of other configurations.
\label{Intro_4}
\end{enumerate}

The equivalence between a quantum field theory and a cellular automaton is not only of conceptual interest. It offers important practical advantages:
\begin{enumi}
\item Concepts of quantum mechanics can be used for an understanding of classical cellular automata, as the density matrix, a change of basis as the Fourier transform, or conserved quantities as momentum that are not directly visible in the updating rule for the automaton.
\label{Intro_i}
\item Cellular automata can be employed for numerical solutions of certain fermionic quantum field theories.
\label{Intro_ii}
\item 
Since probabilistic cellular automata can be viewed as classical generalized Ising models with a positive overall probability distribution many concepts of the functional integral formulation of quantum field theories or statistical mechanics become applicable, as renormalization, perturbation theory, effective theories and dualities.
\label{Intro_iii}
\item A probabilistic cellular automaton is a probabilistic classical computer. This opens prospects to realize features of quantum computing by a probabilistic evolution of "classical" macroscopic yes/no observables, as for artificial neural networks or neuromorphic computing~\cite{CWPC, MOPW, PW}.
\label{Intro_iv}
\end{enumi}

The most important aspect of this work is perhaps the conceptual one. If the dynamics and expectation values of all observables are the same for the probabilistic cellular automaton and the quantum system for fermions, both describe the same physical situation and are equivalent. A quantum field theory realized as a classical statistical generalized Ising model is an example for classical statistical systems realizing quantum mechanics. It constitutes direct evidence that no-go theorems as the ones based on Bell's inequalities do not apply - see ref.~\cite{Wetterich:2020kqi} and references therein for a detailed discussion which assumptions are not satisfied.

\section{Probabilistic cellular automaton\label{sec: 2}}

A (classical) cellular automaton changes the state of a system in a stepwise manner. In our case, at each discrete time step a configuration of classical bits (or fermionic occupation numbers or Ising spins) is updated to a new configuration. The updating is deterministic, with precisely one new configuration following a given configuration. We discuss invertible automata for which the inverse process is also uniquely defined. A probabilistic cellular automaton involves, in addition, a probability distribution over the possible initial configurations, or more general boundary conditions that specify the initial probabilities.

\subsection{Updating rule\label{Updating rule}}

We specify a particular automaton that will be shown to be equivalent to a Thirring type fermionic quantum field theory with two colors. We consider a periodic chain of $M_{x}$ discrete space points~$x$, separated by a distance $\varepsilon$. For every $x$ one has a set of four occupation numbers $n_{\eta a}(x)=(1,0)$, $\eta=1,2=R,L$ for right movers and left movers, and $a=1,2$ for red and green "particles". The total number of bits is $4M_{x}$. Each configuration specifies all occupation numbers, such that there are a total number of $2^{4M_{x}}=N$ configurations. For each time step, the updating proceeds as follows: 

\begin{enumerate}[(1) , itemindent=15pt, leftmargin=20pt, itemsep=3pt, topsep=3pt]
\item First all movers move one position to the right, i.e. $n_{Ra}(t+\varepsilon,x+\varepsilon)=n_{Ra}(t,x)$, and all left movers move one position to the left, $n_{La}(t+\varepsilon,x-\varepsilon)=n_{La}(t,x)$.
\label{UR_1}
\item If a single left mover and a single right mover meet at the same position, the colors are exchanged,
\be\label{A1}
R1+L1\leftrightarrow R2+L2\;,\quad R1+L2\leftrightarrow R2+L1\;.
\ee
\end{enumerate}
The first process can be viewed as propagation, the second as an interaction or scattering. Note that no scattering occurs for sites that are occupied by two right movers with different colors or by two left movers with different colors. Such "bosonic states" propagate without interaction.

\begin{figure}[h]
\centering\includegraphics[width=3in]{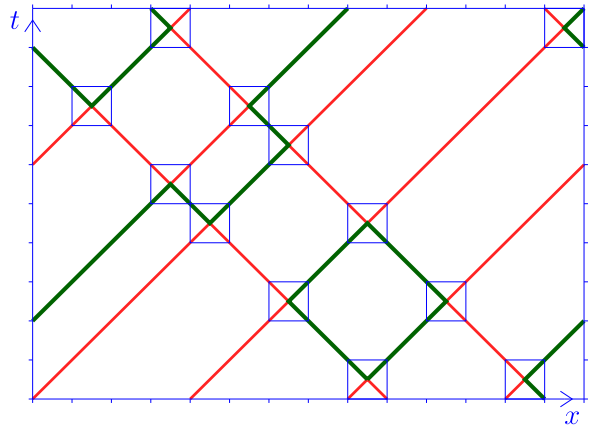}
\vspace*{-5pt}
\caption{Cellular automaton for interacting fermions. Single occupied red (thin) or green (thick) lines scatter at the squares. We have not indicated empty lines, or left- or right moving lines which are doubly occupied by one red and one green particle. These lines are straight without scattering.}\label{Fig. 01}
\end{figure}

We have depicted in Fig.~\ref{Fig. 01} the time evolution for a particular configuration. The discrete points for $x$ and $t$ are in the centers of the squares. We have only indicated the squares for which an interaction takes place - altogether the points $(t,x)$ form a regular quadratic lattice. The evolution of single red and green bits are symbolized by lines - this shows already the analogy with moving particles. 
Empty sites with $n_{Ra}=n_{La}=0$ are left blank. We also have not indicated possible "bosonic lines" with two occupied left movers of different colors, or similar doubly occupied right movers. The corresponding "double lines" would simply be straight lines. They do not affect the propagation of single particle lines.
\noindent

\subsection{\textit{Initial} condition\label{subsec: Initial condition}}

For a deterministic cellular automaton the initial state at some initial time $t_{\textup{in}}$ is given by precisely one specific configuration $\overline{\rho}$. This configuration is propagated by the rules of the automaton to any later time $t$, such that the configuration $\tau$ at $t$ is uniquely determined. This is ideal deterministic classical computing. A convenient description uses an $N$-component real unit vector $q_\textup{in}$ with components $q_{\rho}(t_\textup{in})$. The initial state is specified by $q_{\rho}(t_\textup{in})=\delta_{\rho,\overline{\rho}}$, such that only the $\overline{\rho}\:\!$-component of $q(t_\textup{in})$ differs from zero.

For a probabilistic cellular automaton the initial condition specifies a probability $p_{\rho}(t_\textup{in})$ for every possible initial configuration $\rho$. The probability distribution obeys the standard laws of probability theory, 
$p_{\rho}(t_\textup{in})\geq 0, \sum_{\rho}p_{\rho}(t_\textup{in})=1$. Each configuration $\overline{\rho}$ propagates by the deterministic rules of the automaton to a specific configuration $\tau(t,\overline{\rho})$ at later $t$. The probability to find the configuration $\tau$ at $t$,  $p_{\tau}(t)$, is precisely the probability $p_{\overline{\rho}}(t_\textup{in})$ of the initial configuration from which it originated,
\be\label{PC01}
p_{\tau}(t)=p_{\overline{\rho}(\tau)}(t_\textup{in})\;.
\ee
The updating rule specifies for a basic time step $\varepsilon$ the configuration $\tau\big{(}t+\varepsilon,\rho(t)\big{)}$ for every configuration $\rho(t)$. Together with the initial probability distribution $p_{\rho}(t_{\textup{in}})$ it defines the probabilistic cellular automaton. 

\subsection{Wave function\label{subsec: Wave Function}}

For every $t$ we specify the probability distribution by a wave function $q(t)$,
\be\label{PC02}
p_{\tau}(t)=(q_{\tau}(t))^{2}\;.
\ee
The positivity of the probabilities is guaranteed, and the normalization requires that $q(t)$ is a unit vector 
\be\label{PC03}
q_{\tau}(t)q_{\tau}(t)=1\;.
\ee

The evolution law for the wave function can be written in terms of the step evolution operator $\widehat{S}(t)$ by matrix multiplication
\be\label{PC04}
q(t+\varepsilon)=\widehat{S}(t)q(t)\,,\quad q_{\tau}(t+\varepsilon)=\widehat{S}_{\tau\rho}(t)q_{\rho}(t)\;.
\ee
Indeed, with
\be\label{PC05}
\widehat{S}_{\tau\rho}(t)=\delta_{\tau, \overline{\tau}(\rho)}=\delta_{\overline{\rho}(\tau),\rho}\;,
\ee
the step evolution operator differs from zero only if the configuration $\tau$ at $t+\varepsilon$ equals the configuration $\overline{\tau}(\rho)$ associated to the configuration $\rho$ at $t$ by the rule of the automaton. 
The orthogonal matrix~$\widehat{S}$ has in each row and column precisely one element equal to one, while all others equal zero. It may be called a "unique jump operator".

Eq.~\eqref{PC04} implies
\be\label{PC06}
q_{\tau}(t+\varepsilon)=q_{\overline{\rho}(\tau)}(t)\,,\quad
p_{\tau}(t+\varepsilon)=p_{\overline{\rho}(\tau)}(t)\;,
\ee
thus producing the rule for the probabilistic cellular automaton. Following the evolution from initial time $t_{\textup{in}}$ for a sequence  of time steps yields eq.~\eqref{PC01}. The deterministic cellular automaton is a special case with a sharp initial wave function.

The vector $q(t)$ is the wave function of quantum mechanics in a real representation. (Any complex wave function has an associated real representation with twice the number of components). 
The normalization of the wave functions is guaranteed by the relation~\eqref{PC03}. It is conserved by the evolution ~\eqref{PC04} since $\widehat{S}$ is an orthogonal matrix, such that the length or norm of $q(t+\varepsilon)$ is the same as the one of $q(t)$. Also the relation between the wave function or probability amplitude $q_{\tau}(t)$ and the probabilities $p_{\tau}(t)$ in eq.~\eqref{PC02} is the same as for quantum mechanics. 

\subsection{Step evolution operator\label{subs: Step evolution operator}}

The step evolution operator~$\widehat{S}$ is a central quantity for the present paper. It plays the same role as the unitary evolution operator in quantum mechanics, applied to time differences $\Delta t=\varepsilon$. The particularity for a cellular automaton is that $\widehat{S}$ is a real unique jump operator. Seen in a more general context of classical statistics for which the coordinates $t$ and $x$ are treated on an equal footing, the step evolution operator corresponds to the transfer matrix~\cite{BAX, FUC} in a particular normalization~\cite{CWIT, Wetterich:2020kqi}.

In our case we can write the step evolution operator as a matrix product of a free or kinetic operator $\widehat{S}_{\textup{free}}$ and an interaction operator $\widehat{S}_{\textup{int}}$ , 
\be\label{A2}
\widehat{S}=\widehat{S}_{\textup{int}}\,\widehat{S}_{\textup{free}}\;.
\ee
The operator $\widehat{S}_{\textup{free}}$ propagates all right (left) movers one position in $x$ to the right (left). The interaction operator is a direct product of local operators at each $x$,
\be\label{A3}
\widehat{S}_{\textup{int}}=\widehat{S}_{i}(x_{\textup{in}})\otimes \widehat{S}_{i} (x_{\textup{in}}+\varepsilon)\otimes \widehat{S}_{i}(x_{\textup{in}}+2\varepsilon)\otimes\dots
\ee
Each factor is a $16\times 16$ matrix, involving the $16$ local configurations of the four occupation numbers $(n_{R1},  n_{R2}, n_{L1}, n_{L2})$. It exchanges the configurations $(1\, 0\,0 \, 1)\leftrightarrow (0\, 1\,1 \, 0)$ and $(1\, 0\,1 \, 0)\leftrightarrow (0\, 1\,0 \, 1)$, and leaves all other configurations invariant.

\subsection{Annihilation and creation operators\label{subsec: Annihilation and creation operators}}

One can express $\widehat{S}_{\textup{free}}$ and $\widehat{S}_{i}(x)$ in terms of annihilation and creation operators $a_{\gamma}(x)$, $a_{\gamma}^{\dagger}(x)$, with $\gamma=(\eta, a)$ taking four values. These operators obey the anticommutation relations for fermions
\be\label{A3a}
\{a_{\gamma}^{\dagger}(x),a_{\delta}(y)\}=\delta_{\gamma\delta}\delta_{xy}\ ,\quad \{a_{\gamma}(x),a_{\delta}(y)\}=\{a_{\gamma}^{\dagger}(x),a_{\delta}^{\dagger}(y)\}=0\;.
\ee
One finds for the interaction (the proof will be given elsewhere)
\be\label{A4}
\widehat{S}_{i}(x)=\exp\Big{\{}\frac{i\pi}{2}\big{[}a_{\textup{R}1}^{\dagger}(x)a_{\textup{R}2}(x)-a_{\textup{R}2}^{\dagger}(x)a_{\textup{R}1}(x)\big{]}\big{[}a_{\textup{L}1}^{\dagger}(x)a_{\textup{L}2}(x)-a_{\textup{L}2}^{\dagger}(x)a_{\textup{L}1}(x)\big{]}\Big{\}}\;.
\ee
The kinetic part decomposes into a direct product of independent factors for the four species,
\be\label{A5}
\widehat{S}_{\textup{free}}=\widehat{S}_{1}^{(\textup{R})}\otimes\widehat{S}_{2}^{(\textup{R})}\otimes\widehat{S}_{1}^{(\textup{L})}\otimes\widehat{S}_{2}^{(\textup{L})}\;,
\ee
which obey
\be\label{A6}
\widehat{S}_{a}^{(\textup{R,L})}=N\bigg{[}\exp\Big{\{}\sum_{x}a^{\dagger}(x\pm\varepsilon)\big{[}a(x)-a(x\pm\varepsilon)\big{]}\Big{\}}\bigg{]}\;.
\ee
Here we have omitted the species labels $(\eta,a)$ for the annihilation and creation operators. The different signs distinguish between right movers~$(+)$ and left movers~$(-)$. The ordering operation $N$ moves in each term in the expansion of the exponential all creation operators to the left (with appropriate minus signs). These identities are pure operator identities and do as such not involve any particular fermionic interpretation of the cellular automaton.

\subsection{Hamiltonian and continuum limit}

We can cast the discrete evolution of the wave function for the probabilistic cellular automaton into the form of a continuous Schrödinger equation for quantum mechanics. For this purpose we define the hermitian Hamiltonian $H$ by the relation
\be\label{H1}
\widehat{S}=\exp\big{(}-i\varepsilon H\big{)}\;.
\ee
With the unitary evolution operator~$U(t_{1},t_{2})$ the continuous time-evolution of the wave function
\be\label{H3}
q(t_{2})=U(t_{2},t_{1})q(t_{1})\;,\quad U(t_{1},t_{2})=\exp\big{(}-i(t_{1}-t_{2})H\big{)}\;,
\ee
obeys the Schrödinger equation
\be\label{H4}
i\partial_{t}q=Hq\;.
\ee
For $t_{ \textup{in}}+m\varepsilon$, the solution of this Schrödinger equation coincides with the wave function of the cellular automaton.

One can interpret the probabilistic automaton as a quantum system with time-dependent Hamiltonian, given by $2H_{\textup{free}}$ for $t_{\textup{in}}\leq t\leq t+\varepsilon/2$, and $2H_{\textup{int}}$ for $t+\varepsilon/2\leq t\leq t+\varepsilon$, and continued alternation as $t$ progresses. The interaction Hamiltonian can be read off directly from eq.~\eqref{A4}. We could also modify the Hamiltonian and use the time-independent Hamiltonian
\be\label{H5}
H=H_{\textup{free}}+H_{\textup{int}}+\Delta H\;,
\ee
with $\Delta H$ defined by the relation
\be\label{H6}
e^{-i\varepsilon(H_{\textup{free}}+H_{\textup{int}}+\Delta H)}=e^{-i\varepsilon H_{\textup{int}}}e^{-i\varepsilon H_{\textup{free}}}\;.
\ee

The continuum limit can formally be taken as $\varepsilon\to0$. More precisely, $\varepsilon$ has to be small as compared to a typical time scale for the variation of the wave function. 
It is at this point that the probabilistic nature of the automaton enters crucially. A deterministic cellular automaton has a sharp wave function with every particle placed precisely at a position~$x_{i}$. The wave function changes discountinuously with time, and the continuum limit is not valid in this case. In contrast, a wave function that is smooth in $x$ has also a smooth evolution in $t$.
The modification $\Delta H$ arises from the non-vanishing commutator of $H_{\textup{free}}$ and $H_{\textup{int}}$
\be\label{H7}
\Delta H=\mathcal{O}\big{(}\varepsilon[H_{\textup{int}},H_{\textup{free}}]\big{)}\;.
\ee
It vanishes in the continuum limit $\varepsilon\to0$. Taking also the continuum limit in the space direction, we can omit the ordering operator $N$ in eq.~\eqref{A6}. The ordering plays only a role for contributions~$\sim\big{\lbrace}a^{\dagger}(x), a(y)\big{\rbrace}$ with $y$ precisely equal to $x$. A smooth wave function does not resolve space differences of the order~$\varepsilon$. In order to maintain the orthogonality of the step evolution operator without the ordering one antisymmetrizes the exponent in eq.~\eqref{A6}. Taking further
$\sum_{x}=\varepsilon^{-1}\int\textup{d}x\ ,\ \ a(x+\varepsilon)-a(x-\varepsilon)=2\varepsilon\partial_{x}a(x)$, one finds the continuum limit for the Hamiltonian by comparison of eqs.~\eqref{A4},~\eqref{A6} with eq.~\eqref{H1},
\begin{align}\label{H8}
H_{\textup{free}}=&\frac{i}{\varepsilon}\int\textup{d}x\sum_{a}\Big{\lbrace}a_{La}^{\dagger}(x)\partial_{x}a_{La}(x)-a_{Ra}^{\dagger}(x)\partial_{x}a_{Ra}(x)\Big{\rbrace}\;,\nn\\
H_{\textup{int}}=&-\frac{\pi}{2\varepsilon^{2}}\int\textup{d}x\big{[}a_{\textup{R}1}^{\dagger}a_{\textup{R}2}-a_{\textup{R}2}^{\dagger}a_{\textup{R}1}\big{]}\big{[}a_{\textup{L}1}^{\dagger}a_{\textup{L}2}-a_{\textup{L}2}^{\dagger}a_{\textup{L}1}\big{]}\;.
\end{align}
The factors of $\varepsilon^{-1}$ provide for the correct dimension. They can be absorbed by a rescaling of $a$, $a^{\dagger}$ that leads to the continuum anticommutation relation where $\delta_{xy}$ is replaced by $\delta(x-y)$.

The continuum limit of the evolution of the cellular automaton describes a multi-fermion quantum system in a rather standard way. The evolution with the continuum Hamiltonian $H=H_{\textup{free}}+H_{\textup{int}}$ is unitary since $H^{\dagger}=H$. The step evolution operator~\eqref{H1} does not remain, however, a unique jump operator. The continuum limit corresponds to a type of coarse graining. On this coarse grained level the evolution is probabilistic, in contrast to the deterministic evolution of the microscopic cellular automaton. The continuum step evolution operator~\eqref{H1} does not map a given bit configuration into precisely one new bit configuration. For the free part the displacement by~$\pm\varepsilon$ is only realized in average.

\section{Step evolution operator for fermionic quantum field theories\label{sec: 3}}

We have found a description of the time-evolution of a probabilistic cellular automaton in terms of a Schrödinger equation for a multi-fermion system. The central quantity has been the step evolution operator. In order to deepen the understanding of this fermionic picture we will next discuss this issue in terms of a Grassmann functional integral for a fermionic quantum field theory. This will exhibit a sketch of the key features of a general map between Ising spins and Grassmann variables~\cite{CWFGI,CWFCS}. The central ingredient for the equivalence of a probabilistic cellular automaton with a fermionic quantum field theory is that they share the same step evolution operator and therefore the same dynamics. This can be accompanied by a general map of operators for observables~\cite{CWFGI}. The Grassmann functional integral will reveal the Lorentz symmety of the continuum limit of our model. We will proceed next to the general construction of the step evolution operator for a fermionic functional integral on discrete space-time points.

\subsection{Grassmann functional integral\label{subsec: Grassmann functional integral}}

Consider a Grassmann functional integral
\begin{equation}\label{01}
Z=\int \mathcal{D}\psi \exp(-S[\psi])=\int\mathcal{D}\psi w[\psi]\;,\quad S=\sum_{t}\mathcal{L}(t)\;.
\end{equation}
For $\mathcal{L}(t)$ involving only even powers of anticommuting   Grassmann variables~$\psi$ the weight functional $w[\psi]$ can be written as a product of commuting time local factors $\tilde{\mathcal{K}}(t)$ , 
\be\label{04}
w[\psi]=\exp(-S[\psi])=\!\prod_{t}\tilde{\K}(t)\, , \, \; \tilde{\K}(t)=\exp\lbrace-\mathcal{L}(t)\rbrace\;.
\ee
We consider models for which each local factor depends on two sets of Grassmann variables
$\psi_{\alpha}(t+\tilde{\varepsilon})=\psi_{\gamma}(t+\tilde{\varepsilon}, x)$ and $\psi_{\beta}(t)=\psi_{\delta}(t, y)$ at neighboring $t+\tilde{\varepsilon}$ and $t$. We do not impose space-locality at this stage and leave the range of $x$, $y$ free for the moment.

An element of the local Grassmann algebra at $t$ is a linear combination of Grassmann basis functions, 
\be\label{05}
g(t)=q_{\tau}(t)g_{\tau}(t)\; ,
\ee
where the basis functions $g_{\tau}(t)=g_{\tau}[\psi(t)]$ are products of Grassmann variables $\psi_{\alpha}(t)$. A suitable set of basis functions is
\be\label{06}
g_{\tau}(t)=\tilde{s}_{\tau}\prod_{\alpha=1}^{M}\tilde{a}_{\alpha}\;,
\ee
with $\tilde{a}_{\alpha}=1$ or $\tilde{a}_{\alpha}=\psi_{\alpha}$, and $\tilde{s}_{\tau}=\pm 1$ some conveniently chosen signs. For $\alpha=1\dots M$ there a $2^{M}$ basis functions, $\tau=1\dots N , N=2^{M}$. For the convenience of manipulating signs we also define (no sum over $\tau$)
\be\label{07}
g_{\tau}'=\varepsilon_{\tau}g_{\tau}(t)\,,\quad \varepsilon_{\tau}=(-1)^{\tfrac{m_{\tau}(m_{\tau}-1)}{2}}\;,
\ee
with $m_{\tau}$ the number of $\psi$ - factors in $g_{\tau}$. We observe the identity
\be\label{08}
\exp(\psi_{\alpha}\varphi_{\alpha})=\prod_{\alpha}(1+\psi_{\alpha}\varphi_{\alpha})
= \sum_{\tau}\varepsilon_{\tau}g_{\tau}(\psi)g_{\tau}(\varphi)
=\sum_{\tau}g_{\tau}(\psi)g_{\tau}'(\varphi)\;.
\ee

\subsection{Step evolution operator\label{subsec: Step evolution operator}}

For odd $t=t_{\textup{in}}+(2m+1)\tilde{\varepsilon}$ we define the "transfer matrix" $\widehat{T}_{\tau\rho}(t)$ by the double expansion
\be\label{09}
\tilde{\K}(t)=g_{\tau}(t+\tilde{\varepsilon})\widehat{T}_{\tau\rho}(t)g_{\rho}'(t)\;.
\ee
Adding a constant to $\mathcal{L}(t)$ multiplies $\widehat{T}_{\tau\rho}(t)$ by a constant factor. We can use this freedom to normalize $\widehat{T}_{\tau\rho}(t)$ such that its largest eigenvalues obey $|\lambda_{i}|=1$.
Here "largest" is defined by the largest absolute size, and there may be more than a single largest eigenvalue. With this normalization the transfer matrix becomes the "step evolution operator" $\widehat{S}_{\tau\rho}(t)$. We implicitly assume in the following a suitable normalization of $\mathcal{L}(t)$ such that 
\be\label{10}
\tilde{\K}(t)=g_{\tau}(t+\tilde{\varepsilon})\widehat{S}_{\tau\rho}(t)g_{\rho}'(t)\;.
\ee

Due to the modulo two properties of Grassmann functional integrals~\cite{CWFGI} it is convenient to define the step evolution operator for even $t=t_{\textup{in}}+2m\tilde{\varepsilon}$ by an expansion in conjugate basis functions, 
\be\label{11}
\tilde{\K}(t)=\overline{g}_{\tau}'(t+\tilde{\varepsilon})\widehat{S}_{\tau\rho}(t)\overline{g}_{\rho}(t)\;.
\ee
The conjugate basis functions are defined by the relation
\be\label{12}
\int\mathcal{D}\psi \overline{g}_{\tau}(\psi)g_{\rho}(\psi)=\delta_{\tau\rho}\;,\quad
\overline{g}_{\tau}'=\varepsilon_{\tau}'\overline{g}_{\tau}\;,\quad 
\varepsilon_{\tau}'=(-1)^{m_{\tau}}\varepsilon_{\tau}\eta_{M}(-1)^{Mm_{\tau}}\;.
\ee
For the Grassmann integral at $t$ we order $\int\mathcal{D}\psi=\int\textup{d}\psi_{4}\textup{d}\psi_{3}\textup{d}\psi_{2}\textup{d}\psi_{1}$, with $\psi_{\gamma}=\psi_{\gamma}(t)$.
Up to signs the map from $g_{\tau}$ to $\overline{g}_{\tau}$ exchanges factors of one and $\psi_{\alpha}$ in eq.~\eqref{06}. 
With $\eta_{M}=1$ for $M=0,1 \bmod 4$, that we assume in the following, and $\eta_{M}=-1$ for $M=2,3 \mod 4$ we observe the relations
\be\label{16}
\exp(\psi_{\alpha}\varphi_{\alpha})=\sum_{\tau}g_{\tau}'(\psi)\overline{g}_{\tau}(\varphi)\;,\quad
\int\mathcal{D}\psi g_{\tau}'(\psi)\overline{g}_{\rho}'(\psi)=\eta_{M}\delta_{\tau\rho}\;.
\ee

Consider next the product of two neighboring local factors. For $t$ odd one has
\be\label{21}
\tilde{\mathcal{K}}(t+\tilde{\varepsilon})\tilde{\mathcal{K}}(t)=\overline{g}_{\tau}'(t+2\tilde{\varepsilon})\widehat{S}_{\tau\alpha}(t+\tilde{\varepsilon})F_{\alpha\beta}(t+\tilde{\varepsilon})\widehat{S}_{\beta\rho}(t)g_{\rho}'(t)\,,
\ee
with
\be\label{22}
F_{\alpha\beta}(t)=\overline{g}_{\alpha}(t)g_{\beta}(t) \,, \quad \int\mathcal{D}\psi(t)F_{\alpha\beta}(t)=\delta_{\alpha\beta}\;.
\ee
Integrating the product~\eqref{21} over the common Grassmann variables $\psi(t+\tilde{\varepsilon})$ results in a matrix multiplication of the step evolution operator
\be\label{23}
\int\mathcal{D}\psi(t+\tilde{\varepsilon})\tilde{\mathcal{K}}(t+\tilde{\varepsilon})\tilde{\mathcal{K}}(t)=\overline{g}_{\tau}'(t+2\tilde{\varepsilon})\bigl(\widehat{S}(t+\tilde{\varepsilon})\widehat{S}(t)\bigr)_{\tau\rho} g_{\rho}'(t)\;.
\ee
Similarly, one finds ($t$ odd)
\be\label{24}
\tilde{\mathcal{K}}(t)\tilde{\mathcal{K}}(t-\tilde{\varepsilon})=g_{\tau}(t+\tilde{\varepsilon})\widehat{S}_{\tau\alpha}(t)\tilde{F}_{\alpha\beta}(t)\widehat{S}_{\beta\rho}(t-\tilde{\varepsilon})\overline{g}_{\rho}(t-\tilde{\varepsilon})
\ee
with
\be\label{25}
\tilde{F}_{\alpha\beta}(t)=g_{\alpha}'(t)\overline{g}_{\beta}'(t)\,,\quad \int\mathcal{D}\psi(t)\tilde{F}_{\alpha\beta}(t)=\delta_{\alpha\beta}\;.
\ee
Again, integrating the intermediate Grassmann variable yields a matrix product
\be\label{26}
\int\mathcal{D}\psi(t)\tilde{\mathcal{K}}(t)\tilde{\mathcal{K}}(t-\tilde{\varepsilon})=g_{\tau}(t+\tilde{\varepsilon})\bigl(\widehat{S}(t)\widehat{S}(t-\tilde{\varepsilon})\bigr)_{\tau\rho} \overline{g}_{\rho}(t-\tilde{\varepsilon})\,.
\ee

The product structure can be extended to longer chains of neighboring local factors. Taking the initial time $t_\textup{in}$ as even one can express the partition function by a chain of ordered matrix products of step evolution operators
\be\label{27}
Z=\int\mathcal{D}\psi(t_{\textup{f}})\mathcal{D}\psi(t_\textup{in})g_{\tau}(t_{\textup{f}})\bigl(\widehat{S}(t_{\textup{f}}-\tilde{\varepsilon}) \dots \,
\widehat{S}(t_\textup{in}+1)\widehat{S}(t_\textup{in})\bigr)_{\tau\rho}\overline{g}_{\rho}(t_\textup{in})\;.
\ee
We can write  eq.~\eqref{27} as a matrix trace with a boundary matrix~$\widehat{\mathcal{B}}$,
\be\label{28}
Z= \tr \bigl{\lbrace} \widehat{S}(t_{\textup{f}}-\tilde{\varepsilon}) \dots \widehat{S}(t_\textup{in})\widehat{\mathcal{B}}\,\bigr{\rbrace}\,.
\ee
More general boundary conditions can be implemented by a general form of $\widehat{\mathcal{B}}(t_{\textup{in}}, t_{\textup{f}})$.

\subsection{Propagating and interacting fermions\label{subsec: Propagating and interacting fermions}}
\medskip

The cellular automata for free fermionic quantum field theories are rather simple \cite{Wetterich:2011dt, CWQFT}. What is new in the present work is the construction of cellular automata for fermionic models with interactions. In the present section we pursue systematically the concept of alternating step evolution operators for the propagation and the interaction. This guarantees a unitary evolution by the simple property that each one of the steps is a unique jump operation. The procedure ressembles somewhat the construction of the Feynman path integral by an alternating sequence of momentum and position eigenstates.

For a two-dimensional system we can define the "right transport factor"
\be\label{33}
\tilde{\mathcal{K}}_{R}(t)=\exp\Bigl{\lbrace}\sum_{x}\psi_{\gamma}(t+\tilde{\varepsilon}, x+\varepsilon)\psi_\gamma(t, x)\Bigr{\rbrace}\,.
\ee
The corresponding step evolution operator is a unique jump operator that maps any given state $\rho$ at $t$ to precisely one state $\tau=\overline{\tau}(\rho) $ at $t+\tilde{\varepsilon}$, as specified by eq.~\eqref{PC05}. The left transport factor~$\tilde{\mathcal{K}}_{L}(t)$ obtains by a replacing $\varepsilon\to -\varepsilon$ in eq.~\eqref{33}. The part $\widehat{S}_{\textup{free}}$ of the cellular automaton of the preceding section is realized for
\be\label{41}
\mathcal{L}_\textup{free}(t)=-\sum_{x}\bigl{\lbrace}\psi_{R, a}(t+\tilde{\varepsilon}, x+\varepsilon)\psi_{R,  a}(t,x)
+\psi_{L, a}(t+\tilde{\varepsilon}, x-\varepsilon)\psi_{L,  a}(t,x)\bigr{\rbrace}\;.
\ee

In order to introduce interactions we investigate conditional jumps. We take examples where at every position $x$ the jump is independent of the configurations of occupation numbers at all other positions $x'\neq x$. In this case the local factor $\tilde{\mathcal{K}}(t)$ factorizes into a product of independent factors
\be\label{42}
\mathcal{K}_{\textup{int}}(t)=\prod_{x}\tilde{\mathcal{K}}_{i}(t,x)\,,
\ee
where $\tilde{\mathcal{K}}_{i}(t,x)$ involves only the two sets of Grassmann variables $\psi_{\gamma}(t+\tilde{\varepsilon},x)$ and $\psi_{\gamma}(t,x)$ at the given position $x$. Accordingly, the step evolution operator is a direct product 
\be \label{43}
 \widehat{S}_\textup{int}=\widehat{S}_{i}(x=1)\otimes\widehat{S}_{i}(x=2)\dots\otimes\widehat{S}_{i}(x=M_{x})\;.
\ee
Each factor $\widehat{S}_{i}(x)$ acts only on the configurations of occupation numbers at $x$. For $\gamma=1\dots 4$ each factor $\widehat{S}_{i}(x)$ is a $16\times 16$ matrix and we can discuss each factor $ \tilde{\mathcal{K}}_{i}(t,x)$ or $\widehat{S}_{i}(t,x)$ separately. 

For the conditional jump we first consider the following: Under the condition that precisely two particles are present, namely one left mover and one right mover with different colors, the colors are exchanged. This corresponds to a switch of occupation numbers $(1,0,0,1)\leftrightarrow(0,1,1,0)$.
All other states remain invariant. This process describes the two-particle scatterings $R1+L2\rightarrow R2+L1$ in eq.~\eqref{A1}.
If a third particle is present, no scattering occurs.
The corresponding step evolution operator is a unit matrix except for the sectors of the states $\tau$ with occupation numbers $(1,0,0,1)$ and $(0,1,1,0)$. In this sector the diagonal elements vanish, and one has $\widehat{S}_{(1001),(0110)}=\widehat{S}_{(0110),(1001)}=1$. The two-particle states relevant for our purpose involve the basis functions 
\be\label{50}
g_{(1001)}=-g_{(1001)}'=\psi_{2}\psi_{3}\,,\quad
g_{(0110)}=-g_{(0110)}'=\psi_{1}\psi_{4}\;.
\ee
According to eq.~\eqref{10} the contribution of the two-particle sector to $\tilde{\mathcal{K}}_{i}(x,t)$ reads
\be\label{51}
\begin{split}
\Delta \tilde{\mathcal{K}}_{i}(t,x)=-\psi_{1}(t+\tilde{\varepsilon},x)\psi_{4}(t+\tilde{\varepsilon},x)\psi_{2}(t,x)\psi_{3}(t,x) \\
-\psi_{2}(t+\tilde{\varepsilon},x)\psi_{3}(t+\tilde{\varepsilon},x)\psi_{1}(t,x)\psi_{4}(t,x)\;.
\end{split}
\ee

We have to combine this contribution with the contribution of the unit operator for all other states. For this purpose we first subtract from $\Delta \tilde{\mathcal{K}}_{i}$ the contribution of the unit operator in this particular two-particle sector by defining
\be\label{52}
\tilde{D}(t,x)=-(\psi_{1}'\psi_{4}'-\psi_{2}'\psi_{3}')(\psi_{1}\psi_{4}-\psi_{2}\psi_{3})\;,
\ee
where $\psi_{\gamma}'=\psi_{\gamma}'(t+\tilde{\varepsilon},x), \quad \psi_{\gamma}=\psi_{\gamma}(t,x)$ .
In terms of $\tilde{D}$ we can write the local factor in exponential form $\tilde{\mathcal{K}}_{i}(t,x)=\exp\bigl{\lbrace} -\tilde{\mathcal{L}}_{i}(t,x)\bigr{\rbrace}$ with
\be\label{58}
\tilde{\mathcal{L}}_{i}(t,x)=(-\psi_{\gamma}'\psi_{\gamma}+\tilde{D})(1+\tilde{D})\;,\quad
\tilde{\mathcal{K}}_{i}(t,x)=\exp\lbrace\psi_{\gamma}'\psi_{\gamma}\rbrace-\tilde{D}\;.
\ee
The first term produces the unit matrix, while the second term subtracts the unit matrix in the sector of the states $(1,0,0,1)$ and $(0,1,1,0)$ and replaces it by the exchange of colors.
The construction of the second process $\textup{R}1+\textup{R}2\leftrightarrow \textup{L}1+\textup{L}2$ in eq.~\eqref{A1} proceeds in complete analogy, replacing $\tilde{D}$ by $\overline{D}=\tilde{D}+\tilde{C}$ with
\be\label{83}
\tilde{C}(t,x)=-(\psi_{1}'\psi_{3}'+\psi_{2}'\psi_{4}')(\psi_{1}\psi_{3}+\psi_{2}\psi_{4})\;.
\ee

\subsection{Interacting fermionic quantum field theory\label{subsec: Interacting fermionic quantum field theory}}

For the construction of a quantum field theory for interacting fermions we want to combine the interaction with the propagation of fermions. Similar to the cellular automaton in sect.~\ref{sec: 2} this can be done by the use of a sequence of alternating local factors. We use the free propagation of fermions for $t$ even, and the interaction for $t$ odd. A pair of neighboring local factors reads for even $t$
\be\label{59}
\tilde{\mathcal{K}}(t+\tilde{\varepsilon})\tilde{\mathcal{K}}(t)=\exp\bigl{\lbrace} -\tilde{\mathcal{L}}_\textup{int}(t+\tilde{\varepsilon})\bigr{\rbrace} \exp\bigl{\lbrace} -\tilde{\mathcal{L}}_\textup{free}(t)\bigr{\rbrace}
=\exp\Bigl{\lbrace} -\sum_{x}\bigl[\tilde{\mathcal{L}}_{i}(t+\tilde{\varepsilon}, x)+\tilde{\mathcal{L}}_{f}(t, x)\bigr]\Bigr{\rbrace}\;,
\ee
with $\tilde{\mathcal{L}}_{i}(t+\tilde{\varepsilon}, x)$ given by eq.~\eqref{58} shifted to $t+\tilde{\varepsilon}$, and $\tilde{\mathcal{L}}_{f}(t,x)$  extracted from eq.~\eqref{41},
\be\label{60}
\tilde{\mathcal{L}}_{f}(t,x)=-\psi_{R\alpha}(t+\tilde{\varepsilon}, x+\varepsilon)\psi_{R\alpha}(t,x)
-\psi_{L\alpha}(t+\tilde{\varepsilon}, x-\varepsilon)\psi_{L\alpha}(t,x)\;.
\ee

We could integrate over the variables $\psi(t+\tilde{\varepsilon})$ and obtain with eq.~\eqref{26}
\be\label{61}
\int\mathcal{D}\psi(t+\tilde{\varepsilon})\tilde{\mathcal{K}}(t+\tilde{\varepsilon})\tilde{\mathcal{K}}(t)=g_{\tau}(t+2\tilde{\varepsilon})(\widehat{S}_\textup{int}\widehat{S}_\textup{free})_{\tau\rho}\overline{g}_{\rho}(t)\;.
\ee
Since both $\widehat{S}_\textup{int}$ and $\widehat{S}_\textup{free}$ are unique jump operators, this also holds for the product. The product matrix $
\widehat{S}=\widehat{S}_\textup{int}\widehat{S}_\textup{free}$ describes precisely the step evolution operator of the cellular automaton~\eqref{A2}, which is identical to the one discussed in ref.~\cite{CWPCA}.
Repeating the alternating chain with integration over intermediate Grassmann variables produces matrix chains of $\widehat{S}$. With an implementation of boundary conditions and observables in the fermionic representation, the fermionic model~\eqref{59} is exactly equivalent to the probabilistic cellular automaton of sec.~\ref{sec: 2}.

\subsection{Coarse graining and conjugate Grassmann variables\label{subsec: Coarse graining and conjugate Grassmann variables}}

We could consider the product \eqref{61} as a new local factor $\overline{\mathcal{K}}(t)$ which depends on the Grassmann variables $\psi_{\gamma}(t+2\tilde{\varepsilon},x)$ and $\psi_{\gamma}(x)$. The associated step evolution operators is $\widehat{S}_{\textup{int}}\,\widehat{S}_{\textup{free}}$.
Restricting the observables to even $t$ this defines a "coarse grained" fermionic model.
The relation between $\overline{\mathcal{K}}(t)$ and $\widehat{S}(t)$ is given by $
\overline{\mathcal{K}}(t)=g_{\tau}(t+2\tilde{\varepsilon})\widehat{S}_{\tau\rho}(t)\overline{g}_{\rho}(t)$.
The unit of the time-distance between two neighboring points on the time lattice is arbitrary. For our particular construction a non-interacting particle advances one space unit $\varepsilon$ during two time units $\tilde{\varepsilon}$. We can keep the normalization of the velocity to one by choosing the time difference $\tilde{\varepsilon}$ between neighboring lattice points as $\tilde{\varepsilon}=\varepsilon/2$. 
The disadvantage of integrating out the "intermediate Grassmann variables $\psi_{\gamma}(t+\dfrac{\varepsilon}{2},x)$ consists in the fact that both the Grassmann basis functions $g_{\tau}$ and the conjugate Grassmann basis functions $\overline{g}_{\tau}$ are needed for an extraction of $\widehat{S}(t)$ from $\overline{\mathcal{K}}(t)$, and the simplicity of the exponential factors may not be maintained.

It is often advantageous to keep the simple exponential form of the local factors also for the coarse grained view. This can be done by doubling the number of Grassmann variables at every $t$.
At every $t$ one has then two sets of variables $\psi_{\gamma}(t,x)$ and $\overline{\psi}_{\gamma}(t,x)$, and the functional integration is over $\psi$ and $\overline{\psi}$. In our case we can simply associate $\overline{\psi}(t)$ with $\psi(t+\varepsilon)$ by defining
\be\label{68}
\overline{\psi}_{R,\alpha}(t+\varepsilon,x)=\psi_{R,\alpha}(t+\dfrac{\varepsilon}{2},x)\,, \quad
\overline{\psi}_{L,\alpha}(t+\varepsilon,x)=\psi_{L,\alpha}(t+\dfrac{\varepsilon}{2},x)\;.
\ee
The Grassmann variables $\psi(t+\dfrac{\varepsilon}{2})$ play then a role very similar to the conjugate spinors used, for example, in ref.~\cite{CWFCS}.

With the new definitions we do no longer distinguish between even and odd $t$, and the lattice distance on the coarse grained lattice is the same $\varepsilon
$ in both directions. We will in the following use for $t$ integers $m$ for the sites of the coarse grained lattice, corresponding to even $t$ on the original lattice. The action reads in the coarse grained view
\ba\label{69}
\mathcal{L}(t)=-\sum_{x}&\bigg{\lbrace}\overline{\psi}_{R\alpha}(t+\varepsilon,x+\varepsilon)\psi_{R\alpha}(t,x) +\overline{\psi}_{L\alpha}(t+\varepsilon,x-\varepsilon)\psi_{L\alpha}(t,x)\\
-&\Big[\overline{\psi}_{R\alpha}(t,x)\psi_{R\alpha}(t,x)+\overline{\psi}_{L\alpha}(t,x)\psi_{L\alpha}(t,x)+\overline{D}(x)\Big]    \bigl(1+\overline{D}(x)\bigr)\bigg{\rbrace}\;,\nn
\end{align}
with $\overline{D}(x)$ given by eqs.~\eqref{52} and~\eqref{83} with the identifications
\ba\label{70}
\psi'_{1}&=\psi_{R1}(t,x)\,,\quad \psi_{1}=\overline{\psi}_{R1}(t,x)\,,\quad
\psi'_{2}=\psi_{R2}(t,x)\,,\quad \psi_{2}=\overline{\psi}_{R2}(t,x)\,,\nn\\
\psi'_{3}&=\psi_{L1}(t,x)\,,\quad \psi_{3}=\overline{\psi}_{L1}(t,x)\,,\quad
\psi'_{4}=\psi_{L2}(t,x)\,,\quad \psi_{4}=\overline{\psi}_{L2}(t,x)\;.
\end{align}
The action~\eqref{69} contains the same informations as the action~\eqref{59} since we only have renamed variables.

We may define even and odd sublattices. With $t=m_{t}\varepsilon$~, $x=m_{x}\varepsilon$ and integer $m_{t}$, $m_{x}$ the even (odd) sublattice contains the points with $ m_{t}+m_{x}$ even (odd). The action of our model does not connect the even and the odd sublattice. Since for every step $\varepsilon$ in $t$ the kinetic terms moves the variables either one place to te right or to the left, it does not mix the sublattices. A particle on the even sublattice remains on the even sublattice. The interaction term is local and does not change the situation. In the following we simply omit the odd sublattice and define $\sum_{t,x}$ as a sum over the points of the even sublattice. The factor $\overline{D}(x)$ in eq.~\eqref{69} is evaluated on the even sublattice at $t+\varepsilon$. The sum in eq.~\eqref{69} is over the even sublattice. Eq.~\eqref{69} constitutes the fermionic representation of the cellular automaton discussed in sect.~\ref{sec: 2}. For this discrete formulation no approximations have been made.

\subsection{Continuum limit\label{subsec: continuum limit}}

We define lattice derivatives by
\ba\label{71}
(\partial_{t}+\partial_{x})\psi(t,x)=\dfrac{1}{\varepsilon}\bigl[\psi(t,x)-\psi(t-\varepsilon,x-\varepsilon)\bigr]\,, \nn\\
(\partial_{t}-\partial_{x})\psi(t,x)=\dfrac{1}{\varepsilon}\bigl[\psi(t,x)-\psi(t-\varepsilon,x+\varepsilon)\bigr]\;.
\end{align}
The continuum limit corresponds to $\varepsilon\rightarrow 0$ at fixed distances in $t$ and $x$. For a given distance in time or space the number of intermediate lattice points goes to infinity. Sums are expressed by integrals, 
\be\label{75}
\int dt\int dx=\int_{t,x}=2\varepsilon^{2}\sum_{t,x}
\ee
Here the factor $2\varepsilon^{2}$ accounts for the fact that $\sum_{t,x}$ only sums over the points of the even sublattice.
For a sufficiently smooth wave function the lattice derivatives are replaced by partial derivatives, acting on a continuum of Grassmann variables $\psi_{\gamma}(t,x)$,  $\overline{\psi}_{\gamma}(t,x)$. We also choose a different normalization for the Grassmann variables 
\be\label{76}
\psi(t,x)=\sqrt{2\varepsilon}\psi_{N}(t,x)\;.
\ee
In this way we absorb the factor $(2\varepsilon^{2})^{-1}$ arising from $\sum_{t,x}$. Expressed in terms of $\psi_{N}$ the interaction factor $\overline{D}(x)$ is proportional $4\varepsilon^{2}\psi_{N}^{4}$. The continuum limit is taken at fixed $\psi_{N}$.

The continuum limit simplifies the action considerably. We can omit in eq.~\eqref{69} the factor $\bigl(1+\overline{D}(x)\bigr)$. Furthermore, the difference between $\psi(t,x+\varepsilon)$ and $\psi(t,x)$ can be taken to zero once derivatives have been inserted. We will not write the index $N$ for the renormalized Grassmann variables explicitly. The continuum action takes the simple form of a local fermionic quantum field theory,
\be\label{77}
S=\int_{t,x}\bigl{\lbrace}\overline{\psi}_{R\alpha}(t,x)(\partial_{t}+\partial_{x})\psi_{R\alpha}(t,x)+\overline{\psi}_{L\alpha}(t,x)(\partial_{t}-\partial_{x})\psi_{L\alpha}(t,x)
+2\overline{D}(t,x)\bigr{\rbrace}\;.
\ee
For the local interaction term,
\be\label{78}
\!\!\!\!\overline{D}=-\big{(}\overline{\psi}_{\textup{R}1}\overline{\psi}_{\textup{L}2}-\overline{\psi}_{\textup{R}2}\overline{\psi}_{\textup{L}1}\big{)}\big{(}\psi_{\textup{R}1}\psi_{\textup{L}2}-\psi_{\textup{R}2}\psi_{\textup{L}1}\big{)}
-\big{(}\overline{\psi}_{\textup{R}1}\overline{\psi}_{\textup{L}1}+\overline{\psi}_{\textup{R}2}\overline{\psi}_{\textup{L}2}\big{)}\big{(}\psi_{\textup{R}1}\psi_{\textup{L}1}+\psi_{\textup{R}2}\psi_{\textup{L}2}\big{)},\!\!
\ee
all variables correspond to $\psi_{N}$ and are taken at $(t,x)$.

\subsection{Lorentz symmetry\label{subsec: Lorentz symmetry}}

The continuum limit of our cellular automaton exhibits Lorentz symmetry. This is not very apparent in the updating rule for the automaton. It becomes easily visible for the equivalent Grassmann functional integral. For each color we employ two-component vectors of Grassmann variables
\be\label{LS1}
\psi_{a}=\begin{pmatrix}\psi_{\textup{R}a}\\\psi_{\textup{L}a}\end{pmatrix}\;,\quad \overline{\psi}_{a}=(\overline{\psi}_{\textup{L}a},-\overline{\psi}_{\textup{R}a})\;.
\ee
In this formulation the action takes the form
\be\label{LS2}
S=-\int_{t,x}\Big{\{}\overline{\psi}_{a}\gamma^{\mu}\partial_{\mu}\psi_{a}+\frac{1}{2}\overline{\psi}_{a}\gamma^{\mu}\psi_{a}\overline{\psi}_{b}\gamma_{\mu}\psi_{b}
+\frac{1}{2}\overline{\psi}_{a}\gamma^{\mu}\psi_{b}\varepsilon^{ab}\overline{\psi}_{c}\gamma_{\mu}\psi_{d}\varepsilon^{cd}\Big{\}}\;,
\ee
with the antisymmetric tensor $\varepsilon_{12}=-\varepsilon_{21}=1$.
Here the Dirac matrices are real $2\times 2$ matrices, given by the Pauli matrices
\be\label{90}
\gamma^{0}=-i\tau_{2}\,,\quad \gamma_{1}=\tau_{1}\,,\quad \lbrace\gamma^{\mu},\gamma^{\nu}\rbrace=2\eta^{\mu\nu}\;,
\ee
with Lorentz signature $\eta_{00}=-1$, $\eta_{11}=1$,  $\partial_{0}=\partial_{t}$, $\partial_{1}=\partial_{x}$,
\be\label{91}
\eta_{\mu\nu}=\diag (-1,1)\,,\quad \gamma_{\mu}=\eta_{\mu\nu}\gamma^{\nu}\, ,\quad \overline{\psi}=(\overline{\psi}_{R}, \overline{\psi}_{L})\gamma^{0}\;.
\ee
The Lorentz transformations act on the coordinates in the usual way, and the fermion doublets transform as spinors
\be\label{92}
\delta\psi=-\eta{\Sigma}^{01}\psi\,,\quad \delta\overline{\psi}=\eta\overline{\psi}{\Sigma}^{01}\;,
\ee
with generator
\be\label{93}
{\Sigma}^{01}=\dfrac{1}{4}[\gamma^{0},\gamma^{1}]=\dfrac{1}{2}\tau_{3}.
\ee
The Dirac spinor is composed of two Weyl spinors $\psi_{+}\,,\, \psi_{-}$  that transform independently
\be\label{94}
\psi_{+}=\dfrac{1+\overline{\gamma}}{2}\psi=\begin{pmatrix}
\psi_{R}\\0
\end{pmatrix}
\,,\quad
\psi_{-}=\dfrac{1-\overline{\gamma}}{2}\psi=\begin{pmatrix}
0\\\psi_{L}
\end{pmatrix}\;,\quad\overline{\psi}=(\overline{\psi}_{-},\overline{\psi}_{+})\;,
\ee
where $\overline{\gamma}$ corresponds to $\gamma^{5}$ in four dimensions,
\be\label{95}
\overline{\gamma}=-\gamma^{0}\gamma^{1}=\tau_{3}\,,\quad \lbrace\overline{\gamma},\gamma^{\mu}\rbrace=0\;.
\ee
These Weyl spinors are scaled in opposite directions
\be\label{96}
\delta\psi_{+}=-\dfrac{\eta}{2}\psi_{+}\,,\quad \delta\psi_{-}=\dfrac{\eta}{2}\psi_{-}\;.
\ee
The action~\eqref{LS2} describes a type of Thirring model with two colors.

\section{Discussion\label{sec: Discussion}}

We have discussed a probabilistic cellular automaton and a discretized fermionic quantum field theory. For both models all probabilities follow the same evolution, sharing the same step evolution operator. All expectation values of observables built from classical bits or fermionic occupation numbers are identical.
Both discrete models describe the same physical reality - they are equivalent. The two pictures are related by a general bit-fermion map for both the time-local probabilistic information and the observables~\cite{CWFGI}. 
Our models describe massless fermions in one time and one space dimension. The interaction of the fermions extends previous rather simple settings for the propagation of free fermions. Despite a different formulation of the updating rule the present model is equivalent to the probabilistic cellular automaton discussed in ref.~\cite{PLECH}, where many properties as different ground states, symmetry breaking and topological excitations are described in detail.

With the wave function as a probability amplitude we have employed a concept of quantum mechanics for the description of the classical probabilistic cellular automaton.
Many more quantum concepts are useful for an understanding of the cellular automaton, including the density matrix, operators for observables, the quantum rule for the expectation values of observables, changes of basis as the Fourier transform, the momentum operator or other more general non-commuting operators for observables. In particular, there exists a general complex structure for which complex conjugation is related to the particle-hole transformation. This allows a map to complex quantities. In particular, the density matrix becomes a complex hermitian matrix. Given the limited space of this note we refer for these quantum features in classical statistical systems to refs.~\cite{CWPCA},~\cite{CWIT},~\cite{CWQF}.
The continuum limit leads to further simplifications. The time evolution equation becomes the familiar Schrödinger or von-Neumann equation, with a complex hermitian Hamiltonian~\eqref{H8} consisting of a kinetic and an interaction piece. The kinetic piece involves space-derivatives or the momentum operator. The continuum limit of the Grassmann functional integral reveals the Lorentz symmetry of our model.

On the conceptual side, probabilistic cellular automata are classical statistical systems. They are generalized Ising models for which only a fixed sequence of configurations is allowed between neighboring time-layers, namely those corresponding to the updating rule of the cellular automaton. The forbidden sequences are ensured to have zero probability by defining an action that diverges for any forbidden sequence. The generalized Ising model for the cellular automaton of the present paper can be found in ref.~\cite{PLECH}. The probabilistic aspects of the cellular automaton arise from probabilistic boundary terms, that can be set both at the initial and final time layer. The generalized Ising model is a model on a square lattice with a positive semidefinite overall probability distribution. This classical statistical system can be simulated by standard numerical methods.

One of the advantages of a formulation as a classical statistical system is the possibility of coarse graining Associated methods as functional renormalization permit a continuous extrapolation from the discrete microscopic setting of the cellular automaton to a macroscopic description that typically is continuous. The continuum limit can be viewed as a particular coarse graining. On the coarse grained level the step evolution operator typically does not remain a unique jump operator~\cite{Wetterich:2020kqi}. Columns and rows of $\widehat{S}$ contain then more than one non-zero entity. Elements can be negative, or become complex. On the coarse grained level the evolution becomes probabilistic, in contrast to the deterministic step evolution operator of the microscopic cellular automaton. Nevertheless, for a suitable coarse graning there are subsystems for which the information is not erased for for the evolution between time layers. For such subsystems the step evolution operator is unitary. The same quantum formalism with wave functions etc.  applies on the coarse grained level. This reveals that the particular deterministic evolution of the cellular automaton is not crucial for describing quantum systems as particular classical statistical systems. Quantum mechanics can emerge from classical statistics.

Our two-dimensional model remains still rather simple. If these concepts could be generalized to four dimensions, with interactions giving rise to the standard model of particle physics and gravity, they could provide for a classical statistical description of our quantum world.

\enlargethispage{20pt}

\competing{'The author(s) declare that they have no competing interests’.}


\end{document}